\documentclass[12pt]{iopart}

\usepackage{graphics}
\usepackage{bm}
\usepackage{color}

\def\PRA{{\it Phys.~Rev.~A} }

\def\JPB{{\it J.~Phys.~B} }
\def\PRL{{\it Phys.~Rev.~Lett.} }
\def\RMP{{\it Rev.~Mod.~Phys.} }
\def\JCP{{\it J.~Chem.~Phys.} }

\newcommand{\myscaleboxa}[1]{\scalebox{0.6}[0.6]{#1}}
\newcommand{\myscaleboxb}[1]{\scalebox{0.6}[0.4]{#1}}
\newcommand{\myscaleboxc}[1]{\scalebox{0.4}[0.4]{#1}}

\begin{document}

\title{Theory of high-order harmonic generation from molecules by intense laser pulses}

\author{Anh-Thu Le$^1$, R. Della Picca$^2$, P.~D. Fainstein$^2$, D. A. Telnov$^3$,
M. Lein$^4$ and C.~D. Lin$^1$}

\address{$^1$ J. R. Macdonald Laboratory, Department of Physics,
Kansas State University, Manhattan, Kansas 66506, USA}

\address{$^2$ Centro Atomico Bariloche, Comisi\'on Nacional de Energ\'ia Atomica,
Avda E. Bustillo 9500, 8400 Bariloche, Argentina}

\address{$^3$ Department of Physics, St.~Petersburg State University,
St.~Petersburg 198504, Russia}

\address{$^4$ Institut f\"ur Physik and Center for Interdisciplinary Nanostructure
Science and Technology, University of Kassel,
Heinrich-Plett-Stra{\ss}e 40, 34132 Kassel, Germany}


\begin{abstract}
  We show that high-order harmonics generated from molecules by
  intense laser pulses can be expressed as the product of a
  returning electron wave packet and the photo-recombination cross
  section  (PRCS) where the electron wave packet can be obtained from
  simple strong-field approximation (SFA) or from a companion atomic
  target. Using these wave packets but replacing the PRCS
  obtained from SFA or from the atomic target by the accurate PRCS
  from molecules, the resulting HHG spectra are shown to agree well
  with the benchmark results from direct numerical solution of the
time-dependent Schr\"odinger equation,  for the case of H$_2^+$ in
laser fields. The result illustrates that these powerful
theoretical tools can be used for obtaining high-order harmonic
spectra from molecules. More importantly, the results imply that
the PRCS extracted from laser-induced HHG spectra can be used for
time-resolved dynamic chemical imaging of transient molecules with
temporal resolutions down to a few femtoseconds.

\end{abstract}

\submitto{\JPB}
\maketitle


High-order harmonic generation (HHG) is one of the most studied
nonlinear phenomena in intense laser-matter interaction
\cite{brabec00}. HHG is most easily understood using the
three-step model \cite{kulander,corkum,lewenstein} -- first the
electron is released by tunnel ionization; second, it is
accelerated by the oscillating electric field of the laser and
later driven back to the target ion; and third, the electron
recombines with the ion to emit a high energy photon. A
semiclassical formulation of the three-step model based on the
strong-field approximation (SFA) is given by Lewenstein {\it et
al} \cite{lewenstein}. In this model (often called the Lewenstein
model), the liberated continuum electron experiences the full
effect from the laser field, but the effect of the target
potential is neglected. Despite this limitation, the Lewenstein
model has been widely used for studying HHG from atoms and
molecules, and for characterizing the nature of the harmonics
generated. Since the continuum electron needs to come back to
revisit the parent ion in order to emit radiation, the neglect of
the electron-ion interaction in the SFA model is rather
questionable. In the past years, various efforts have been made to
improve upon the SFA model, i.e., by including the Coulomb
distortion \cite{ivanov,kaminski,ciappina,smirnova}, or by using
Ehrenfest theorem \cite{gordon}. All of these improvements can
still be considered insufficient since they fail to include
accurately the scattering of the continuum electrons with the
parent ions.

With moderate effort, direct numerical solution of the
time-dependent Schr\"odinger equation (TDSE) for atomic targets
can be accurately carried out, at least within the single active
electron (SAE) approximation. The TDSE methods are computationally
more demanding for molecular targets. Even for simplest diatomic
molecules, one has to solve a three-dimensional time-dependent
Schr\"odinger equation provided the molecular axis has an
arbitrary orientation with respect to the polarization of the
laser field. Such calculations require fast CPU and large memory
computers and have been accomplished only recently
\cite{lein03,kamta,telnov07}. In order to compare with
experimental results, thus most of the existing calculations for
HHG from molecules were carried out using the SFA model
\cite{zhou,atle06,hoang}. Based on the experience from high
harmonics generated by atomic targets, such SFA model is not
expected to offer accurate predictions. The lack of a reliable
theory for describing HHG from molecules has prevented the
possible exploitation of molecular structure using infrared laser
pulses. This is unfortunate since it has been well recognized that
infrared lasers have the potential for probing time-resolved
molecular dynamics, with temporal resolutions down to sub-ten
femtoseconds.

 In view of the difficulty in solving the TDSE as a practical theoretical tool  for
 predicting the nonlinear interactions between molecules with
 intense lasers, recently Le {\it et al} \cite{atle08} and Morishita {\it et
 al} \cite{toru08} have developed an alternative approach for
 calculating HHG, and have tested the model for atomic targets. The theory is based on the
 three-step model but without the approximations that lead to the SFA (or the Lewenstein model).
 The HHG is viewed as resulting from the photo-recombination of
 the returning electrons generated by the laser pulses. The HHG yield is
 expressed as the product of a returning wave packet with the
 photo-recombination cross section (PRCS). The shape of the wave packet has been
 shown to be largely independent of the targets for the same laser
 pulse and can thus be obtained from a reference atom or from the SFA. Thus there are
 two ways to obtain HHG spectra without
 actually solving the TDSE. The first method is called SW-SFA
 model where the HHG spectra are obtained by replacing the dipole
 matrix element using scattering waves for the continuum electrons
 instead of the plane waves used in the SFA, but retaining the
 electron wave packet from the SFA. The other method is to compare
 the HHG spectra with another reference atom where   the HHG
 spectra have been obtained by accurate calculations or by
 experiment. If the accurate PRCS for both the reference atom and
 the target are known, then the HHG spectra can be
 obtained from the HHG spectra of the other atom. The validity of
 the two methods have been established in Le {\it et al} \cite{atle08} and Morishita {\it et
 al} \cite{toru08} for atomic systems where the results from the
 models are compared to accurate results from the TDSE. While the
 model has been tested only for atoms so far, there is a general
 belief that the same models would  apply to molecular targets.
 In this paper our goal is to apply the models to the H$_2^+$ molecules
 where accurate HHG spectra have been calculated by solving the
 TDSE \cite{lein03,kamta,telnov07}. The results from this comparison confirm
 that the models indeed work well and the method can be further extended
 as a general tool for calculating HHG from molecular targets.

  First let us be specific about the two methods for obtaining HHG
  spectra. The first method is called the scattering wave based strong-field approximation
(SW-SFA) \cite{atle08}. The  HHG yield $S(\theta,\omega)$ for a
molecule whose internuclear axis makes an alignment angle $\theta$
with respect to the laser polarization direction can be obtained
from the SFA result $S^{SFA}(\theta,\omega)$ by
\begin{equation}
S(\theta,\omega)=\left|\frac{T(\theta,E)}{T^{PWA}(\theta,E)}\right|^2
S^{SFA}(\theta,\omega) \label{SW-SFA}
\end{equation}
where $T$ and $T^{PWA}$ are the {\em exact} transition dipole
matrix element and its plane-wave approximation (PWA),
respectively. Here the electron energy $E$ is related to the
emitted photon energy   $\omega$ by $E=k^2/2=\omega-I_p$, with
$I_p$ being the ionization potential of the target (atomic units
are used throughout the paper unless otherwise indicated). As only
the ratio of the transition dipoles enters Eq.~(1), one can
formulate the SW-SFA model by using the above equation with the
photoionization or its time-reversed process, i.e., the
photo-recombination cross sections. In this paper we refer to the
photoionization process, as it is more widely available
theoretically and experimentally. Note that, one can identify, up
to some factor, $S^{SFA}/|T^{PWA}|^2$  as the flux of the
returning electron, which we will call a ``wave-packet''. The
returning electron that contributes most to the HHG is the one
that propagates along the laser polarization direction. Therefore,
the relevant differential cross section is for ${\bi k}$ that is
parallel to the laser polarization axis, that is, for $\theta_{\bi
k}=0$ and $\pi$. Note that for asymmetric molecules, the two
contributions from $\theta_{\bi k}=0$ and $\pi$ differ, in
general.

Clearly, the formulation of the SW-SFA given in Eq.~(1) is not unique. In fact, one
can also use the ``wave-packet'' from a reference atom
(i.e., atom with the same $I_p$ as for the target). In other words, the HHG yield
can also be written as
\begin{equation}
S(\theta,\omega)=\left|\frac{T(\theta,E)}{T^{ref}(E)}\right|^2
S^{ref}(\omega) \label{scaledH}
\end{equation}
where the superscript ``{\it ref}'' refers to the reference atom.
In the following, we will check the validity of both models by
comparing their predictions against the benchmark data. For the
reference atom, we will use a scaled hydrogen with the effective
nuclear charge chosen such that it has the same $1s$ binding
energy as H$_2^+$. The advantage of using  the scaled hydrogen as
a reference atom is that one can accurately and efficiently
calculate both the HHG yield by solving the TDSE numerically, and
the exact photo-recombination cross section analytically.
Experimentally one can also replace the scaled atomic hydrogen
with an atomic target with comparable ionization energy.

In this paper we will focus only on the HHG component in the
parallel polarization direction. The theoretical methods used to
generate the data for the present study have all been given
previously. The ``benchmark'' HHG spectra of H$_2^+$ are calculated
by solving the TDSE using the methods given in
\cite{lein03,telnov07}. The calculation of the transition
dipole matrix elements are described in \cite{picca06,picca07}
while the HHG calculated within the SFA is given in
\cite{zhou,atle06,hoang}. In terms of computational effort and
thus the accuracy, the solution of the TDSE is the most demanding.

\begin{figure}
\centering
\mbox{\rotatebox{0}{\myscaleboxb{
\includegraphics{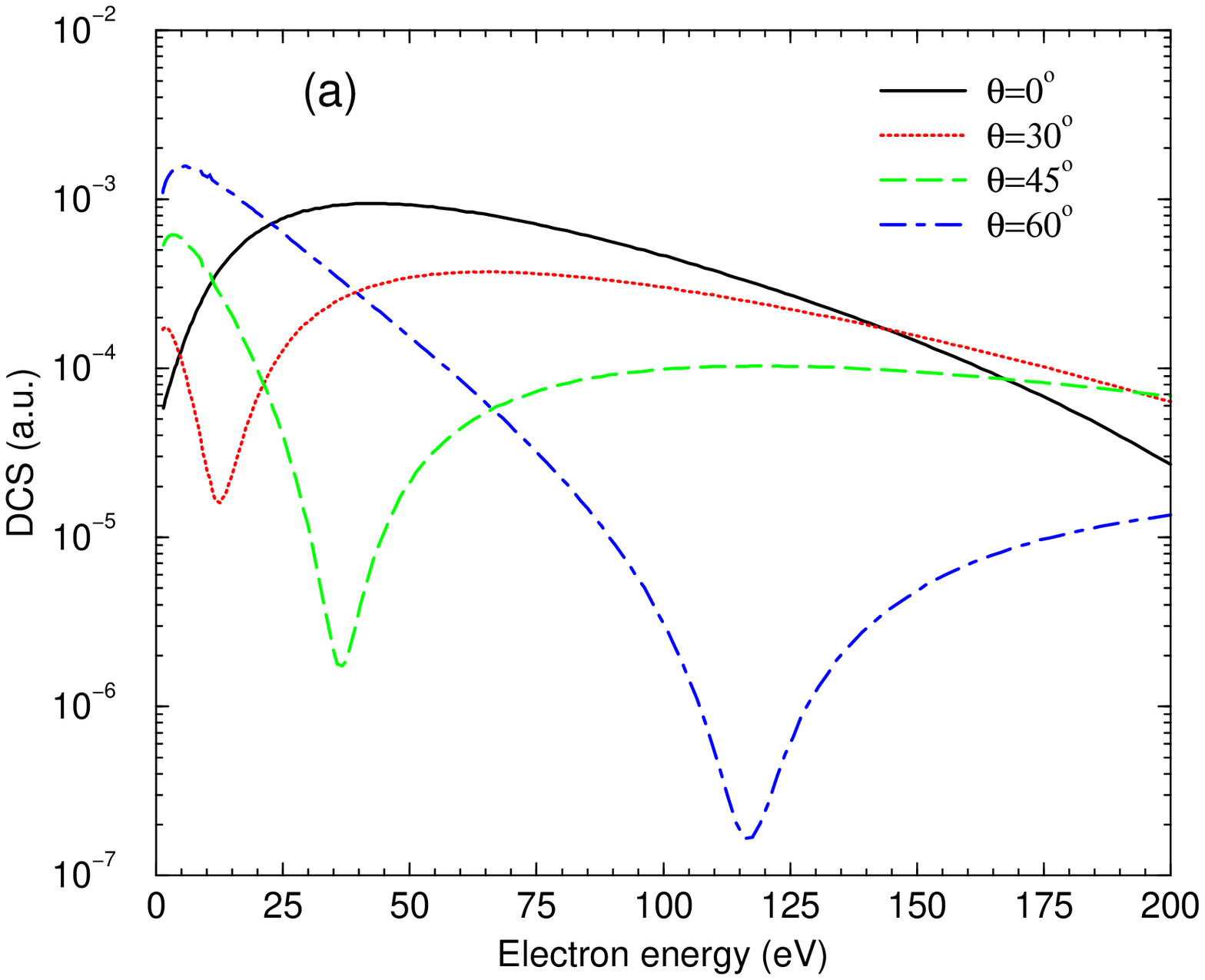}}}}
\mbox{\rotatebox{0}{\myscaleboxb{
\includegraphics{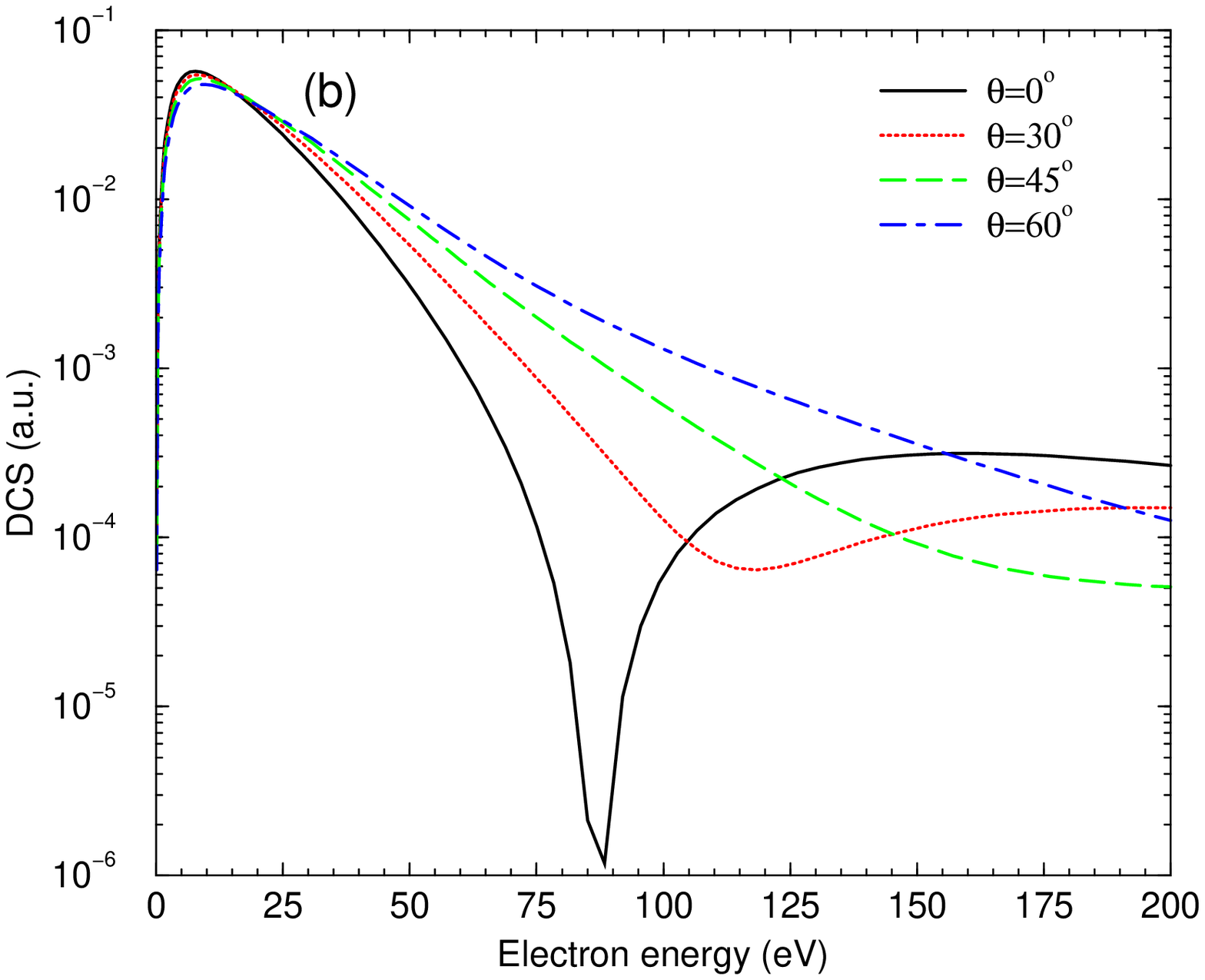}}}}
\caption{Comparison of the photoionization differential cross sections from the
exact (i.e.,with scattering wave) and the plane-wave approximation
calculations, shown in (a) and (b), respectively, for alignment
angles $\theta=0^{\circ}$, $30^{\circ}$, $45^{\circ}$ and
$60^{\circ}$.} \label{fig1}
\end{figure}

In Fig.~1(a) we show the  photoionization differential cross
sections (DCS) (electrons moving along the polarization axis), as
functions of emitted electron energy, for H$_2^+$ at the
equilibrium distance $R=2.0$ a.u., alignment angles
$\theta=0^{\circ}$, $30^{\circ}$, $45^{\circ}$ and $60^{\circ}$.
The data are shown for the energy range up to 200 eV, relevant to
harmonic generation from typical infrared lasers.   The noticeable
features of these curves are the pronounced minima which move to
the higher energies, as the alignment angle increases. The minimum
positions are $12$ eV, $36$ eV and $117$ eV, for
$\theta=30^{\circ}$, $45^{\circ}$, and $60^{\circ}$, respectively.
As we will show later, these minima are responsible for the
interference minima, seen in the HHG spectra. The corresponding
data obtained from the PWA are plotted in Fig.~1(b). Apart from
the clear differences in the shape of these curves in comparison
to the exact data, there are strong discrepancies in the positions
of the minima. For $\theta=30^{\circ}$, the minimum is shifted to
higher energy as much as about $100$ eV. For the two larger
angles, the minima are not even seen within the energy range of
the plot. That is the reason that the minima in the HHG spectra
from SFA are shifted to much higher harmonic orders in comparison
to the exact TDSE calculations, as observed by Kamta and Bandrauk
\cite{kamta} and by Chirila and Lein \cite{chirila}.

\begin{figure}
\centering \mbox{\rotatebox{0}{\myscaleboxc{
\includegraphics{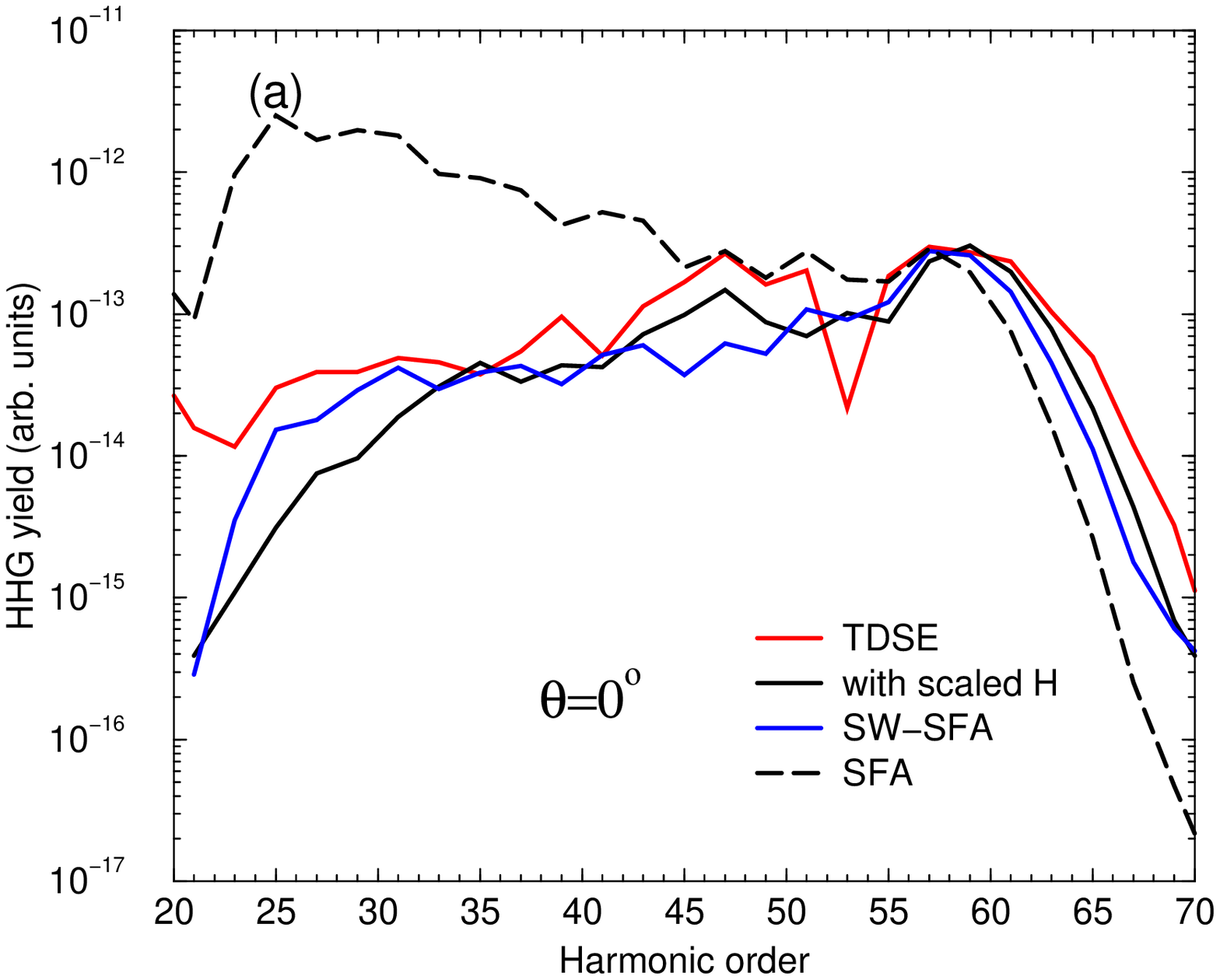}}}}
\mbox{\rotatebox{0}{\myscaleboxc{
\includegraphics{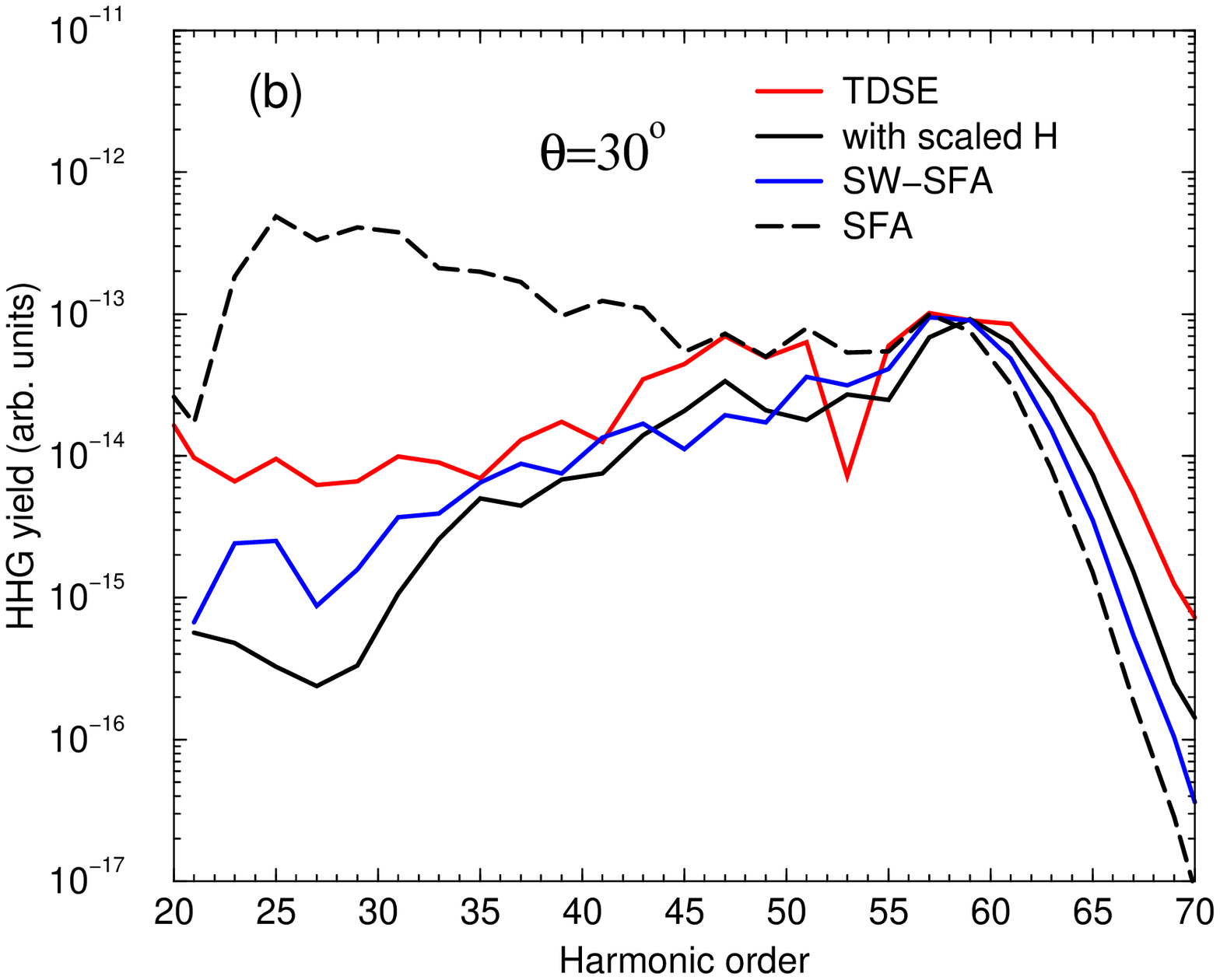}}}}
\mbox{\rotatebox{0}{\myscaleboxc{
\includegraphics{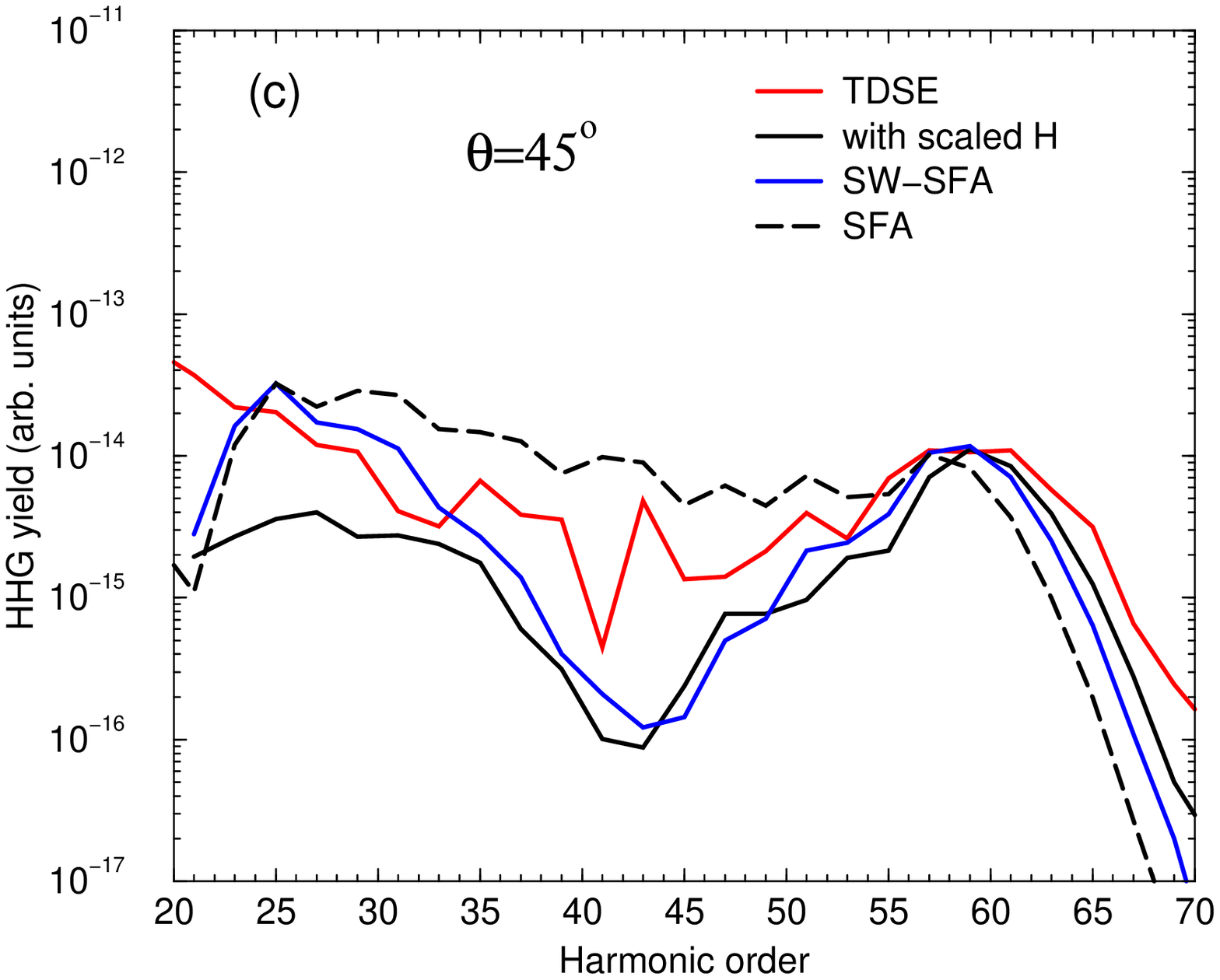}}}}
\mbox{\rotatebox{0}{\myscaleboxc{
\includegraphics{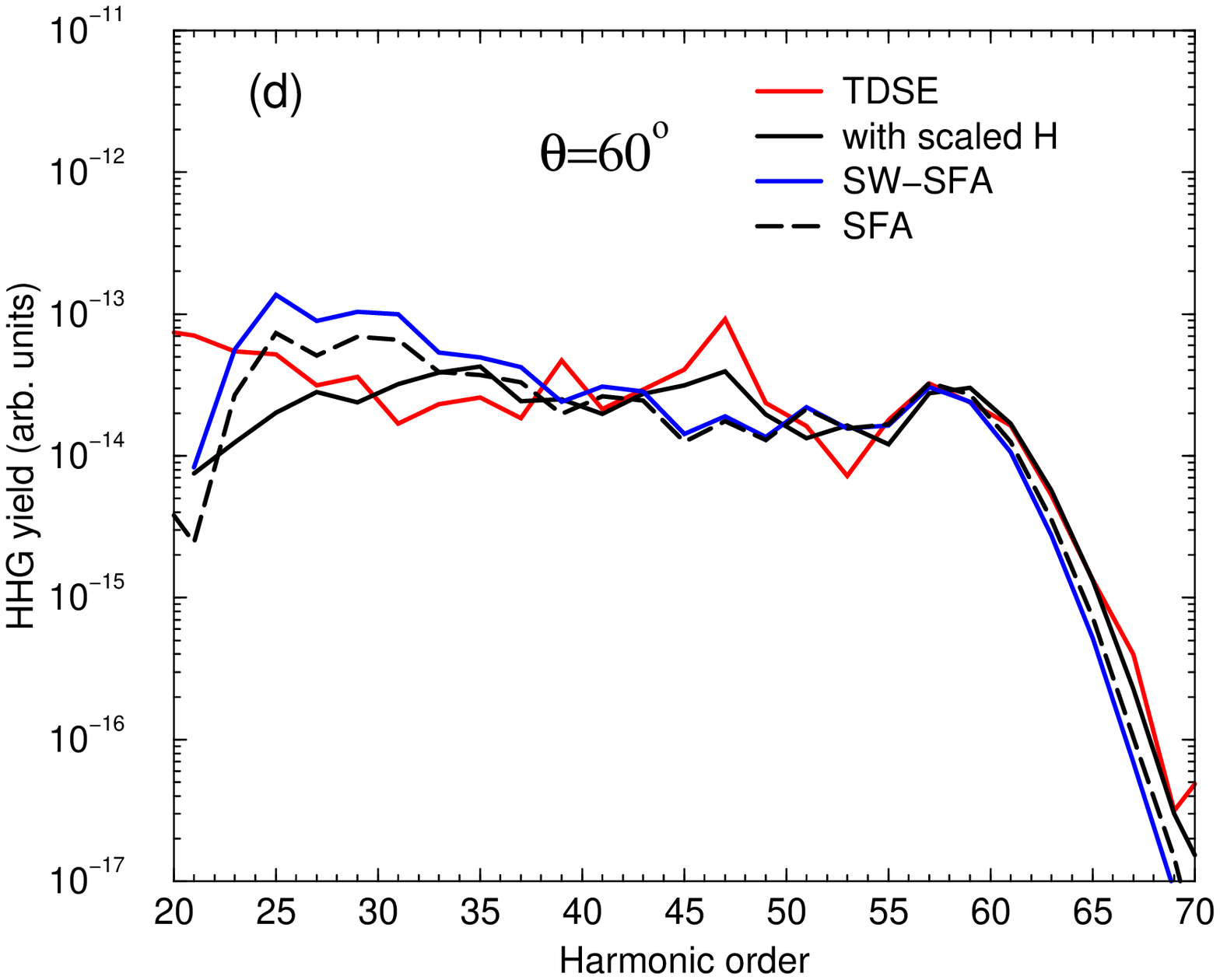}}}}
\caption{Comparison of the HHG spectra obtained from the TDSE (solid red lines),
SFA (dashed lines), SW-SFA (solid blue lines), and with the use of scaled hydrogen
atom (solid black lines) for the alignment angle $\theta=0^{\circ}$, $30^{\circ}$,
$45^{\circ}$ and $60^{\circ}$. A 20-cycle laser with peak intensity of
$3\times 10^{14}$ W/cm$^2$, wavelength of 800 nm is used.} \label{fig2}
\end{figure}

Next we compare the HHG spectra calculated using the different
models with those obtained from the solution of the TDSE.  In
Fig.~2(a-d) we show the TDSE (solid red lines) and  SFA (dashed
black lines) results against those from using SW-SFA (solid blue
lines) and with the wave-packet from scaled hydrogen (solid black
lines), for $R=2.0$ a.u. and $\theta=0^{\circ}$, $30^{\circ}$,
$45^{\circ}$, and $60^{\circ}$. We use laser pulse with peak
intensity of $3\times 10^{14}$ W/cm$^2$, wavelength of 800 nm, and
20-cycle duration with a sine-square envelope. The TDSE results
for H$_2^+$ were obtained by using the method described in
Ref.~\cite{telnov07}. Here we have normalized the data near the
cutoff and only the odd harmonics are shown.

For $\theta=0^{\circ}$ and $30^{\circ}$, one can see that the SFA
results disagree greatly with the TDSE results. The improvements by
the SW-SFA and with the wave-packet from the scaled hydrogen are
quite striking. Indeed, results from both models reproduce quite
well the overall shape of the spectra over the broad range down to
H21, i.e., just above the ionization threshold ($I_p=30$ eV). For
$\theta=30^{\circ}$ the minimum in both model calculations near H27
comes from the dip near $E=12$ eV in the exact DCS [see Fig.~1(a)].
This minimum is consistent with the TDSE data  \cite{telnov07},
although not clearly seen in Fig.~1(b). In fact, this is also
consistent with the earlier results by Lein {\it et al}
\cite{lein03} and Kamta and Bandrauk \cite{kamta}. For
$\theta=45^{\circ}$, the minimum in both of our model calculation is
near H43, also agrees well with the prediction from all the
available TDSE results reported in Refs.~\cite{kamta} (see their
Table II), and \cite{telnov07}. Again, this minimum comes from the
dip near $36$ eV in the exact DCS. We note that the minima from our
model calculations here tend to be more pronounced, compared to the
TDSE results. The remaining discrepancies could be due to the
limitation of our model, or because of the possible lack of
convergence in the TDSE results. Finally, for $\theta=60^{\circ}$,
we notice that all the data, including the SFA, agree quite well
with the TDSE results. This is not entirely surprising since the
shape of the photoionization DCS for this angle from the PWA is
fairly close to that of the exact DCS for the relevant electron
energy range below $70$ eV [see Fig.~1(a) and (b)].

\begin{figure}
\centering \mbox{\rotatebox{0}{\myscaleboxb{
\includegraphics{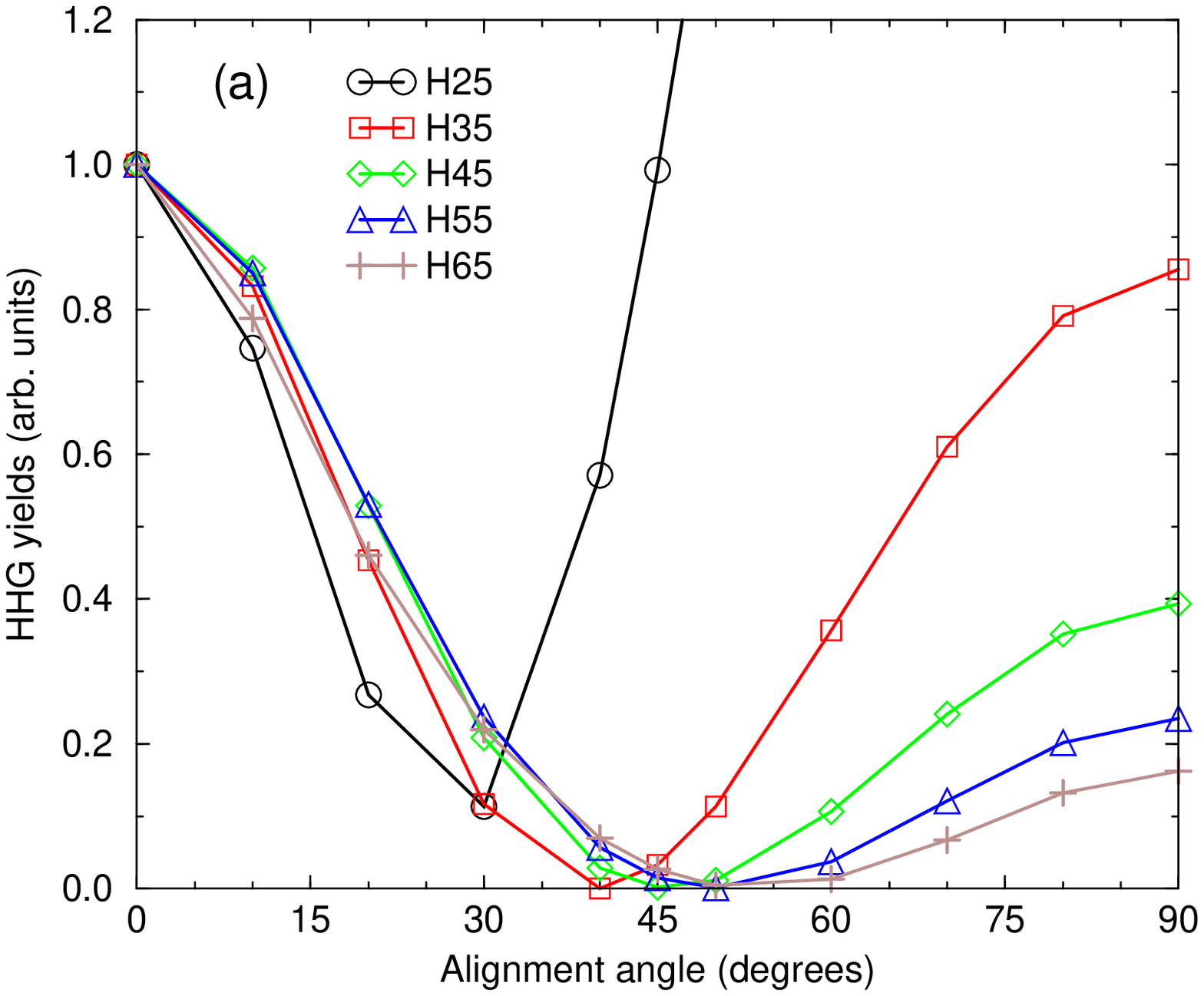}}}}
\centering \mbox{\rotatebox{0}{\myscaleboxb{
\includegraphics{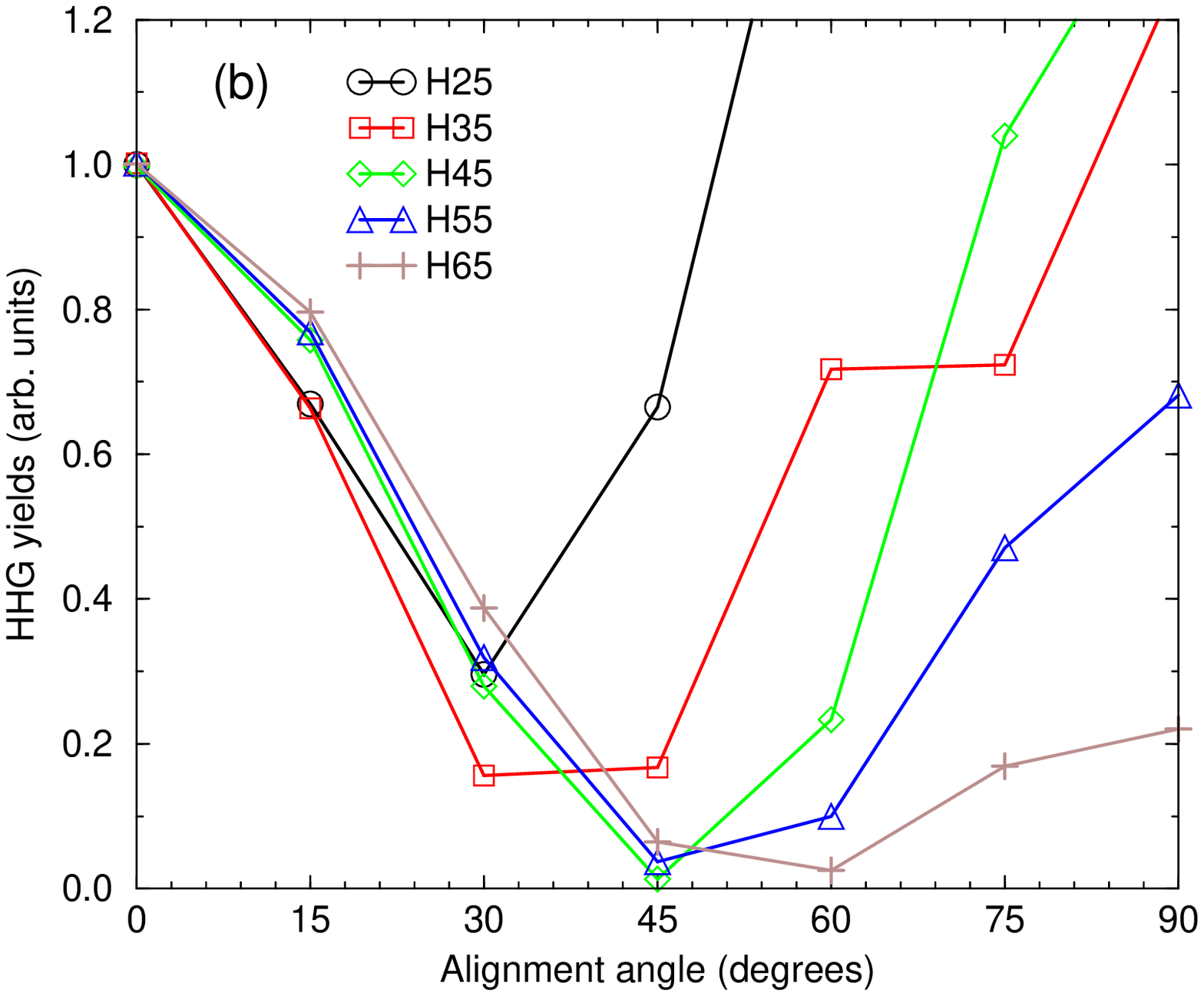}}}}
\caption{Alignment dependence of some selected harmonics from the SW-SFA (a) and the
TDSE (b).} \label{fig3}
\end{figure}

It is important to emphasize that for each new alignment the
returning wave-packet from the SFA is calculated again, whereas
the wave-packet from scaled hydrogen is, of course, alignment
independent. It has been shown \cite{toru08,atle08} that the shape
of the wave-packet largely depends only on the laser field (for
systems with identical ionization potentials). Using scaled
hydrogen, one therefore can get only the {\em shape} of the
wave-packet, as function of energy. To get the absolute magnitude
of the HHG spectra from the latter, one needs to multiply by a
factor, which accounts for the alignment dependent ionization
yield. This can be done, for example, by using the MO-ADK theory
\cite{moadk}.

To better quantify the prediction of the SW-SFA model for
different molecular alignment against the TDSE results, we show in
Fig.~3(a) and (b) the selective harmonics, H25, H35, H45, H55 and
H65, respectively, against the alignment angles. For convenience,
the data were normalized to $1.0$ at $\theta=0^{\circ}$. Again,
the general shape, as well as the minimum position for each
harmonic, are in quite good agreement.

\begin{figure}
\centering \mbox{\rotatebox{0}{\myscaleboxa{
\includegraphics{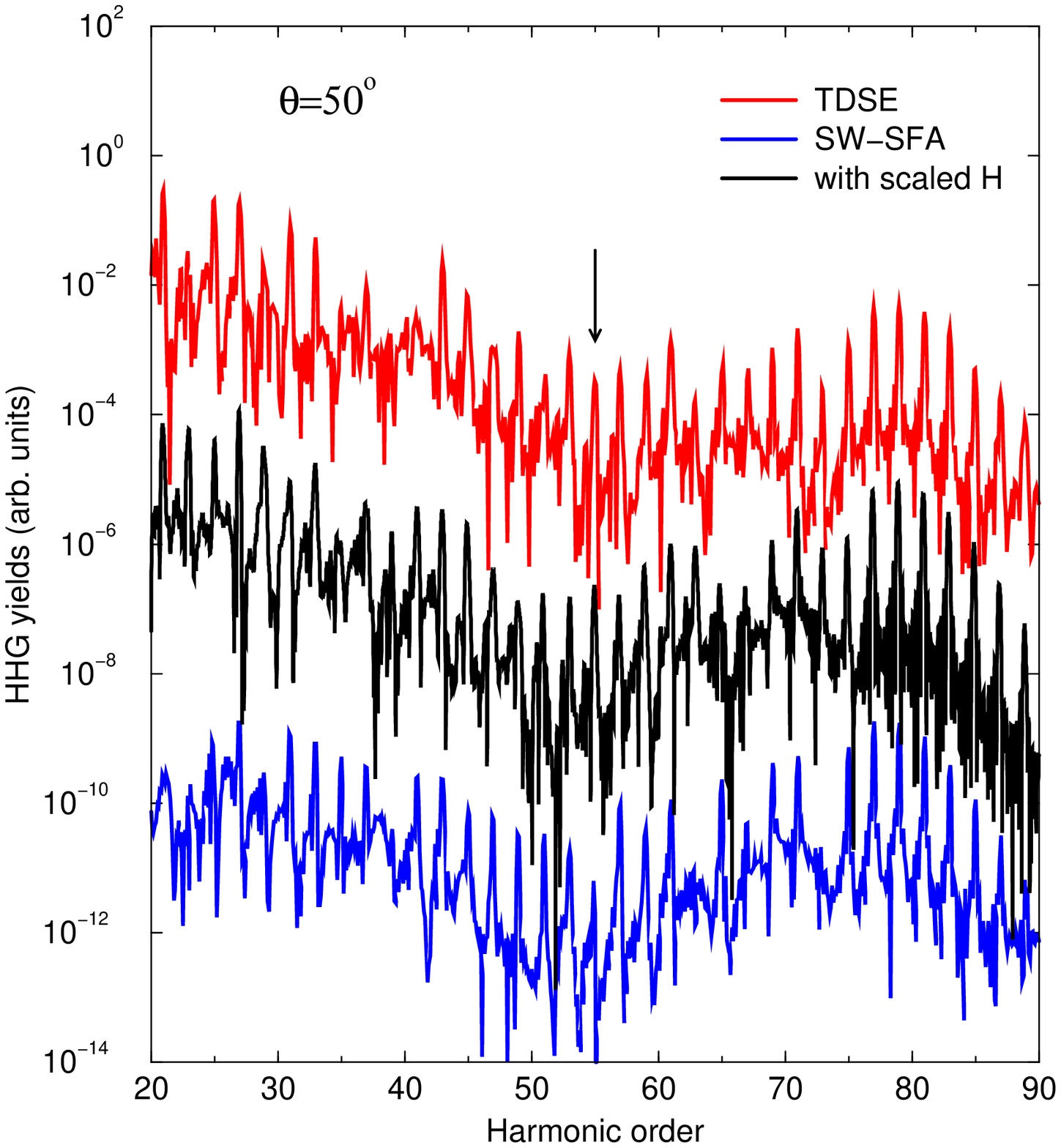}}}}
\caption{Comparison of the HHG spectra obtained from the TDSE (red
line), SW-SFA (blue line), and with the use of the returning
wave-packet from scaled hydrogen (black line), for alignment angle
$\theta=50^{\circ}$. The data have been shifted vertically to show
the details. A laser pulse with peak intensity of $5\times
10^{14}$ W/cm$^2$, wavelength of 780 nm is used. The pulse is
turned on and off over three cycles and kept constant for four
cycles. The arrow indicates the position of the interference
minimum.} \label{fig4}
\end{figure}

We now show detailed comparison of the HHG spectra from different
calculations in Fig.~4 for the alignment angle
$\theta=50^{\circ}$. Here we use 10-cycle laser pulse with peak
intensity of $5\times 10^{14}$ W/cm$^2$ and wavelength of $780$
nm, the same as has been used by Lein {\it et al} \cite{lein03}.
The pulse is turned on and off over three cycles and kept constant
for four cycles. The TDSE results were obtained by the method
described in Ref.~\cite{lein03}. For clarity, we have shifted the
data vertically. Clearly, the SW-SFA data (blue line) are in quite
good agreements with the TDSE results (red line). The results from
our model using the wave-packet calculated by solving the TDSE
equation for scaled hydrogen (black line) agree even better with
the TDSE results. In particular, the minimum near H55, indicated
by an arrow in the figure, is well reproduced by both model
calculations.

\begin{figure}
\centering \mbox{\rotatebox{0}{\myscaleboxa{
\includegraphics{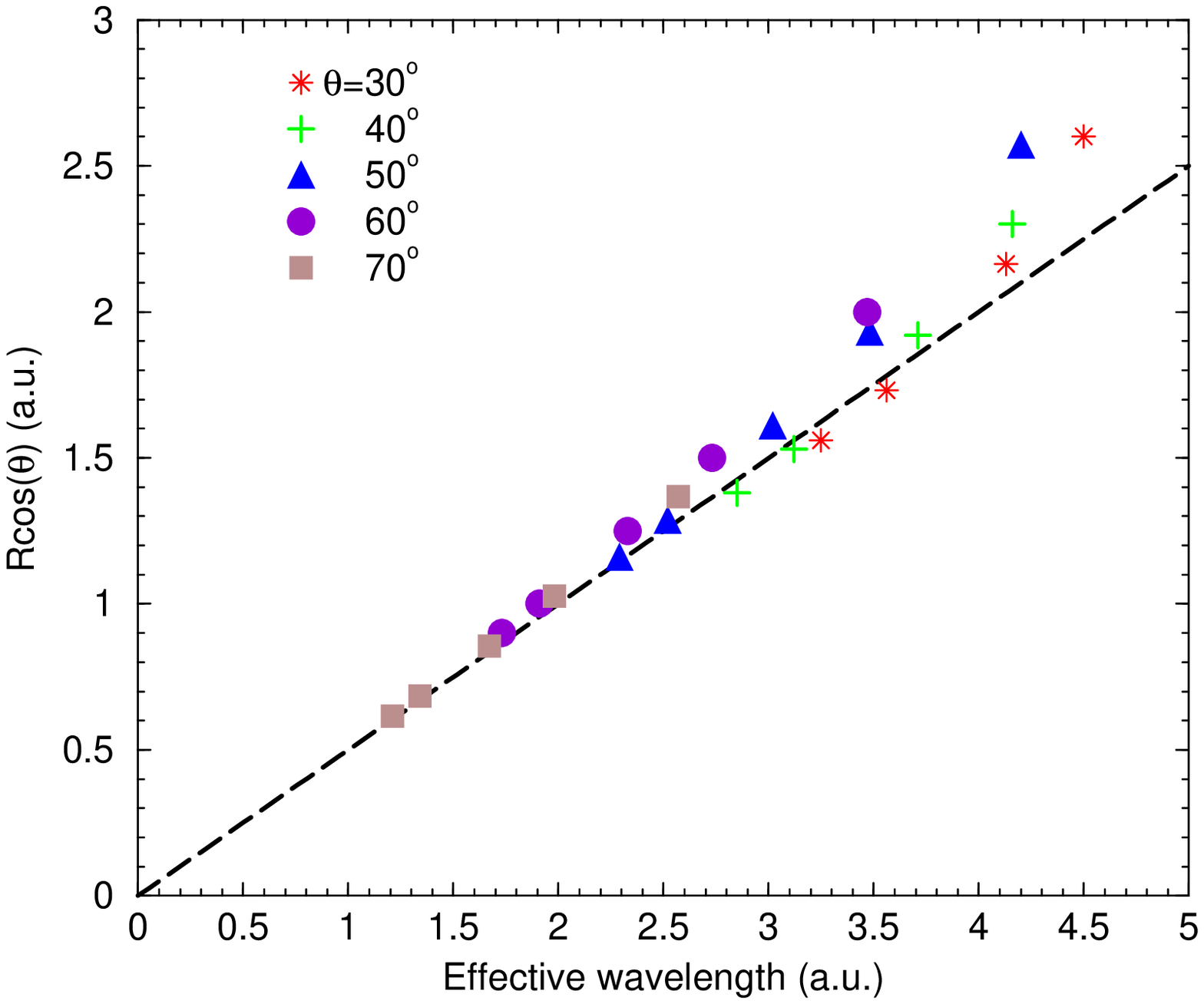}}}}
\caption{Projected internuclear separation vs ``effective''
electron wavelength at the minima in the {\em exact}
photoionization cross sections. Data were obtained with $R=1.8$,
$2.0$, $2.5$, $3.0$, and $4.0$ a.u., and for different alignment
angles as shown in the labels. Note that we have used the
``effective'' wavelength according to Lein {\it et al}
\cite{lein02}, see text.} \label{fig5}
\end{figure}

 In recent years, the ``interference minima'' in the HHG spectra
 from molecular targets have been widely discussed in the
 literature \cite{leinPRL,kanai,vozzi,atle06}. As shown by Lein {\it et al}
\cite{lein02}, the minimum positions in HHG spectra from the TDSE
for H$_2^+$ with $R=1.8$, $2.0$ and $2.5$ a.u. agree quite well
with the two-emitter model \cite{leinPRL,lein02}. Specifically,
the minima satisfy the relation $R\cos(\theta)=\lambda^{eff}/2$,
where $\lambda^{eff}$ is the ``effective'' wavelength of the
continuum electron defined such that the ``effective'' wave vector
is $k^{eff}=\sqrt{2\omega}$, with $\omega$ being the energy of the
emitted photon. In other words, the ``effective'' energy is
shifted by $I_p$ with respect to the usual relation, i.e.,
$k=\sqrt{2(\omega-I_p)}$.

 According to the two models presented here, the minima in HHG are
 attributed to the minima in the photoionization DCS. In Fig.~5
 the projected internuclear
separation $R\cos(\theta)$ is plotted vs the ``effective''
electron wavelength at the minima in the photoionization DCS. The
data were obtained  for $R=1.8$, $2.0$, $2.5$, $3.0$ and $4.0$
a.u. and for alignment angles $\theta=30^{\circ}$, $40^{\circ}$,
$50^{\circ}$ $60^{\circ}$ and $70^{\circ}$. The results are indeed
scattered nicely around $R\cos(\theta)=1/2\lambda^{eff}$, shown as
the dashed line, for small $\lambda^{eff}$ and tend to lie above
it for $\lambda^{eff}>3.5$ a.u. This is consistent with the
results from the numerical solution of the TDSE by Lein {\it et
al} \cite{lein02} [see their Fig.~(3)]. This result demonstrates
that the minima in the HHG spectra from H$_2^+$ are fully and
accurately reproducible by the minima in the photoionization cross
sections, and hence, by the SW-SFA model, for other internuclear
distances as well. We note in this connection that analysis of the
recombination cross sections in relation with the minima positions
in the HHG spectra from H$_2$ and N$_2$ has been reported recently
by Zimmermann {\it et al} \cite{zimmermann}.


In this paper we have shown that harmonic generation from H$_2^+$
molecule can be factored out into the product of a returning
electron wave packet with the photo-recombination cross section. The
wave packet can be extracted from the SFA model or from HHG
generated by atoms with similar ionization potential. Using either
wave packet, but replacing the PRCS of either case by the accurate
PRCS of the molecule, we show that the resulting HHG spectra
compared favorably with those from solving the TDSE. Similar
conclusions have been reached for atomic targets recently
\cite{atle08,toru08}. The models have been tested using different
laser parameters and for different internuclear distances, and we
expect the models to work well for molecules in general. Since
photoionization (or photo-recombination) is a linear process, it is
intrinsically much easier to treat theoretically than the nonlinear
laser-molecule interaction. While accurate calculations of molecular
photoionization DCS are by no means trivial, sophisticated packages
have been developed \cite{gianturco,lucchese,cherepkov} over the
years. Furthermore, for the purpose of obtaining HHG spectra for
molecules in the intense laser pulses, high-precision cross sections
in a narrow energy region such as those measured with synchrotron
radiation are not needed. Instead, moderately accurate results over
a broad range of energy such as those based on the one-electron
model developed by Tonzani \cite{tonzani} is likely adequate.

The significance of the present result is not limited to a
workable theory for HHG from molecules. Most importantly, the
results suggest that it is possible to extract photoionization
cross sections over a broad energy range from the measured HHG
spectra which may then be used to unravel the structure of the
target molecule. Since infrared lasers with sub-ten femtoseconds
are already widely available, high-order harmonics generated from
molecules undergoing transformation may then be used to extract
the structure of the transient molecules at different time delays.
The potential of using HHG from molecules for time-resolved
chemical imaging appears quite promising.

This work was supported in part by Chemical Sciences, Geosciences
and Biosciences Division, Office of Basic Energy Sciences, Office
of Science, U.S. Department of Energy. M.L. acknowledges support
from Deutsche Forschungsgemeinschaft.

\end{document}